\newcommand*{\ARXIV}{}

\ifdefined\ARXIV
    \documentclass{article}
    \usepackage{arxiv}
    \usepackage{moreverb,url}
    \usepackage[hidelinks]{hyperref}
    \usepackage[square,numbers]{natbib}
    \usepackage[pdftex]{graphicx}
    \newcommand{\citecustom}[1]{(\cite{#1})}
    \usepackage{stmaryrd}
\fi

\usepackage{amsmath}
\usepackage{graphicx}
\usepackage{graphbox}
\usepackage{amssymb}
\usepackage{cases}
\usepackage{booktabs}

\def\eg{e.g.,\ }

\def\cf{\textit{cf.}\ }

\newcommand{\CoP}{\ifmmode\mbox{CoP}\else CoP\fi}
\newcommand{\CxP}{\ifmmode\mbox{CxP}\else CxP\fi}
\newcommand{\GC}{\ifmmode\mbox{GC}\else GC\fi}

\usepackage{xcolor}
\usepackage[normalem]{ulem}
\newcommand\soutpars[1]{\let\helpcmd\sout\parhelp#1\par\relax\relax}
\long\def\parhelp#1\par#2\relax{%
  \helpcmd{#1}\ifx\relax#2\else\par\parhelp#2\relax\fi%
 }

	\usepackage{algorithm}
	\usepackage{algorithmicx}
	\usepackage[noend]{algpseudocode}
		\algrenewcommand\algorithmicindent{2.0em}%
		\algnewcommand\Let[2]{\State #1 $\gets$ #2}
		\algnewcommand\AND{\ \textbf{and}\ }
		\algnewcommand\OR{\ \textbf{or} \ }
		\algnewcommand\algorithmicinput{\textbf{Input:}}
		\algnewcommand\Input{\item[\algorithmicinput]}
		\algnewcommand\algorithmiccompute{\textbf{Compute:}}
		\algnewcommand\Compute{\item[\algorithmiccompute]}
		\algnewcommand\algorithmicoutput{\textbf{Output:}}
		\algnewcommand\Output{\item[\algorithmicoutput]}
		\algblockdefx{MRepeat}{EndRepeat}{\textbf{repeat}}{}
		\algnotext{EndRepeat}
\newcommand{\Argmin}[1]{\ensuremath{\mathrm{Arg}\underset{#1}{\mathrm{min}\,}}}

\newcommand{\Avg}[1]{\ensuremath{\underset{#1}{\mathrm{Avg}\,}}}

\newcommand{\titleName}{Decision Support System for an Intelligent Operator of Utility Tunnel Boring Machines}

\newcommand{\abstr}{
In tunnel construction projects, delays induce high costs. Thus, tunnel boring machines (TBM) operators aim for fast advance rates, without safety compromise, a difficult mission in uncertain ground environments. Finding the optimal control parameters based on the TBM sensors’ measurements remains an open research question with large practical relevance. 

In this paper, we propose an intelligent decision support system developed in three steps. First past projects performances are evaluated with an optimality score, taking into account the advance rate and the working pressure safety. Then, a deep learning model learns the mapping between the TBM measurements and this optimality score. Last, in real application, the model provides incremental recommendations to improve the optimality, taking into account the current setting and measurements of the TBM. 

The proposed approach is evaluated on real micro-tunnelling project and demonstrates great promises for future projects.
}
\newcommand{\keywo}{Decision Support System, Intelligent Operator, Utility Tunnel Boring Machines.}
\newcommand{\aknow}{The authors would like to express their gratitude to Herrenknecht AG, Schwanau, Germany, for providing the data and supporting the project with their domain expertise. This work was supported by the Swiss National Science Foundation (SNSF) Grant no. PP00P2-176878.}

\begin{document}
\ifdefined\ARXIV
    \title{\titleName}
    \author{%
        Gabriel Rodriguez Garcia\\
        ETH Z\"urich,\\
        Z\"urich, Switzerland\\
        \And 
        Gabriel Michau\\
        ETH Z\"urich,\\
        Z\"urich, Switzerland\\
        \And
        Herbert H. Einstein\\
        Massachusetts Institute of Technology,\\
        Cambridge, MA 02139-4307, USA\\
        \And
        Olga Fink\\
        ETH Z\"urich,\\
        Z\"urich, Switzerland}
        \subtitle{Preprint}
        \date{08th of January 2021}
        \maketitle
        \begin{abstract}
        \abstr
        \end{abstract}
\fi

\section{Introduction}
\label{sec:intro}

Construction of tunnels has been gaining importance, driven by population growth and urbanization, enabling expansion and enhancement of transportation and underground utility networks. However, tunnelling is subject to large uncertainties  of geology and other construction conditions that contribute to the frequently occurring delays and overruns of budgeted costs in such projects. A good management of risks and uncertainties is, therefore, one of the key requirements of large tunnelling projects. Decision support systems (DSS) are one of the primary tools supporting the decision makers.

To assess and reduce tunnelling risks, several DSS have been developed already several decades ago, most of which were addressing general projects risks prior to tunnel construction. One of the widely applied DSS in the tunneling industry is the Decision Aids for Tunnelling (DAT) that were introduced in 1998 \citet{Einstein1998} and have since been further developed , integrating several additional features (\citet{Einstein2004}, \citet{Christoph2006}, \citet{Sangyoon2016}, \citet{moret2016construction}). The DAT enable quantitative risk estimations in the form of time-cost distributions given geology, resource and construction strategy information. 

Other studies (\citet{Sousa2012}, \citet{Mooney2014}, \citet{Zhao2019}) have conducted research on dynamically updating and predicting the geology ahead of the TBM to decrease the main source of uncertainty and to support decision makers during excavation.

Because of time and budget pressure, the goal of tunnel boring machine (TBM) operators is advancing fast and safely through the tunnel trajectory, achieving high advance rates, while maintaining a high availability of the TBM (since machine breakdowns and unavailability would lead to large delays). advance rates depend on the design, configuration and system state of the TBM but also on the current geological conditions. While TBM have recently been increasingly equipped with numerous sensors to monitor the operation, information on the exact geology is still very scarce and its inference from other measured parameters remains an open research question. Therefore, it cannot be directly integrated in the control. Experienced TBM operators are often able to deduce good control parameter values that enable them to adjust to the geological conditions and achieve good advance rates. However, the level of experience varies among the operators and often, sub-optimal control parameters are used. Finding optimal values for the control parameters based on the information from the TBM system, the operating and the geological conditions remains an open research question with a large practical relevance. 

 A similar challenge as in tunnelling was addressed by \citet{Payette2015} with the goal of supporting operator actions in real-time well drilling for the oil and gas industry. The drilling advisory system (DAS) helps operators to avoid drilling dysfunctions acting as performance limiters and helps to achieve higher advance rates while extending drill bit lifetime \citecustom{Payette2015}. It analyzes data from active control parameters (based on the investigation of drill-off tests) and visually maps optimal parameter settings for the current geologic formation \citecustom{Sanderson2017}. However, drill-off tests are, typically, not part of the common practice in tunneling applications. This approach is, therefore, not directly transferable to TBM operation. 
 
Another approach applied for optimizing drilling parameters for oil wells tackles similar challenges as those addressed in the current research study (\citet{Chandra2019}, \citet{Ryan2016}, \citet{Jiang2016}). In \citet{Chandra2019}, artificial neural networks are applied to predict the advance rate. Heuristic search algorithms (Particle Swarm Optimization or Ant Colony Optimization) are then used to find optimal control parameters in real-time. This approach is different from the one proposed here since it does not consider the scarcity of the data available for training and may require a simulator if abundant data are not available.

To the best of our knowledge, there have been no published studies tackling real-time control parameter optimization with a DSS in the field of tunnel construction risk management. The framework proposed here aims at developing an intelligent decision support system that recommends optimal control parameters in real time to the machine operators (Figure \ref{fig:framework}). The proposed approach only requires the availability of representative TBM operational data. In addition, a measure of credibility is defined to inform the operator to which extent the model suggestions are supported by training samples under similar conditions. This is particularly relevant in the context where operating regimes exhibit a high variability and where some operating regimes may contain only few measured observations. Furthermore, we define a novel and intuitive measure to quantitatively describe optimality of operation for a micro-tunnelling TBM based on advance rate and working pressure.

\section{Framework}
\label{sec:fw}
\begin{figure*}
\includegraphics[width=\textwidth]{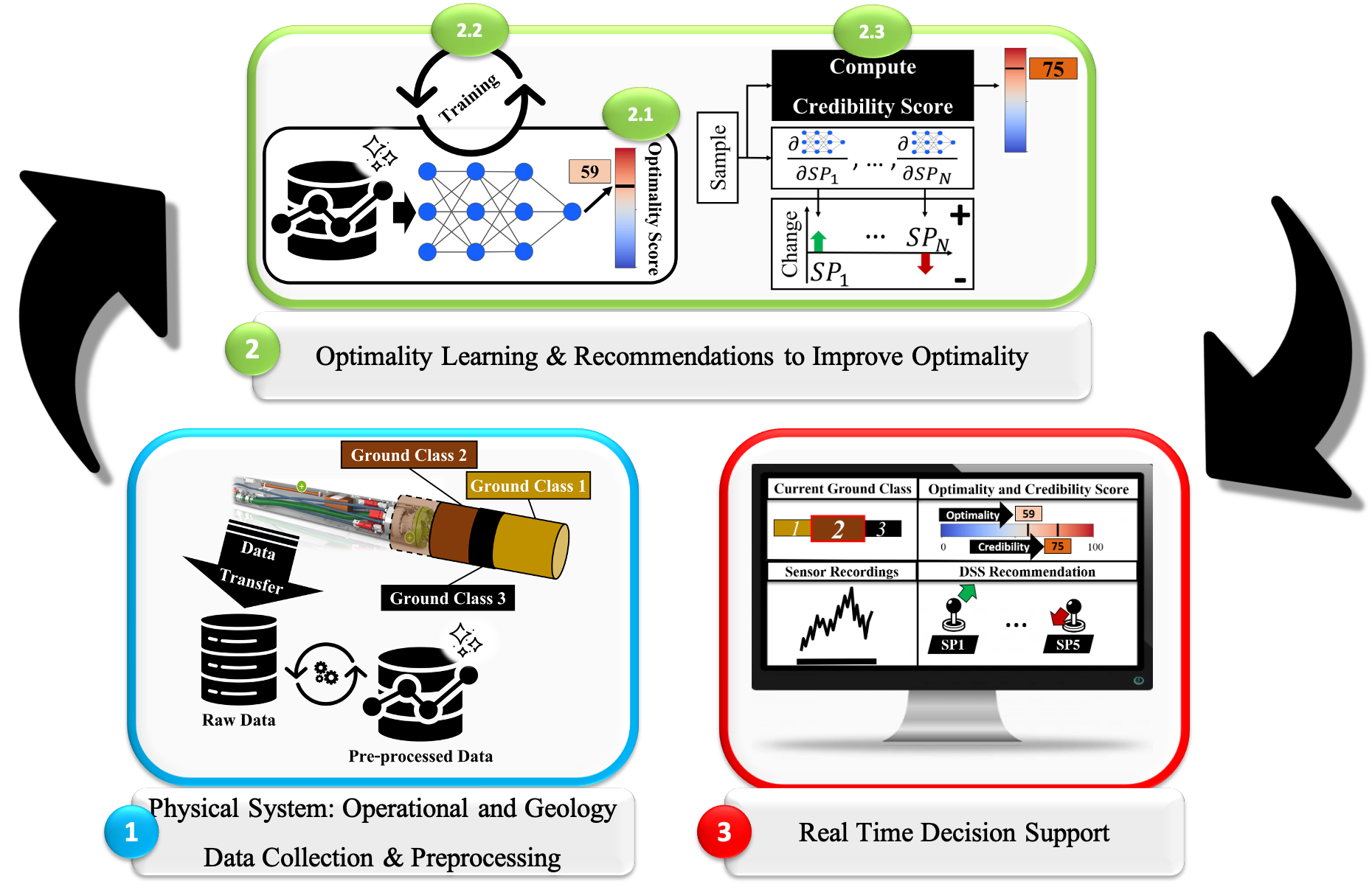}
\caption{\textbf{Framework}\label{fig:framework} Visual representation of the framework, describing how sensor data is leveraged to support TBM operators while drilling. (1) All data sources, including TBM sensor measurements and geological information are pre-processed and cleansed. (2) We train a Deep Learning Feed-forward Neural Network on the available historical data to predict the ``optimality score'' (2.1). The trained model (2.2) can be used during the TBM operation to compute recommendations for control parameters and their credibility in real time (2.3). (3) A dashboard presents these recommendations to the operator.}
\end{figure*}

The proposed framework, as illustrated in Figure~\ref{fig:framework}, consists of three main steps. First, all data sources, including TBM sensor measurements and geological information are pre-processed and cleansed (Fig.~\ref{fig:framework} - 1). Second, we train a Deep Learning Feed-forward Neural Network on the available historical data to predict the ``optimality score'', the measure of the boring efficiency proposed in this work (Fig.~\ref{fig:framework} - 2.1). Third, the trained model (Fig.~\ref{fig:framework} - 2.2) can be used during the TBM operation to compute recommendations for control parameters and their credibility in real time (Fig.~\ref{fig:framework} - 2.3). A dashboard presents these recommendations to the operator (Fig.~\ref{fig:framework} - 3).

\section{Related Works}
\label{sec:related_works}

\noindent\textbf{Decision Aids for Tunnelling (DAT)}

Given the significant uncertainties and their effect on tunnelling projects, research on general tunnelling risk assessment and management, in particular for developing DAT systems. has been abundant (\citet{Einstein2017}, \citet{Mikaeil2019}, \citet{Wang2019}). Existing DAT are advanced tools integrating several features. They have already been in use in large tunnel construction projects for decades, including the Gotthard and Lötschberg Basis Tunnels in Switzerland. The DAT are decision support systems that estimate the distribution of possible costs, time and resource requirements, considering a multitude of inputs such as geologic conditions, available resources and construction strategies. They rely on Monte Carlo simulations of a large number of construction cycles, given possible geology profiles. Subsequently, the time to finish the corresponding tunnels with the user-defined strategy and boundary conditions is calculated for each of the excavation cycles \citecustom{Einstein1998}. 

\citet{Sousa2012} developed a related approach to what is proposed in this paper. This approach is based on Bayesian Networks and comprises two models. The first model predicts the ground class (soil, mixed or rock) based on geology sensitive sensor parameters. The second model infers the construction strategy with the lowest risk (open mode or closed mode excavation for Earth Pressure Balance Machines) based on the predicted ground class. The geology prediction model is updated along the trace using information on already excavated tunnel sections to improve the classification accuracy. Compared to the current research study, the DAT and their extensions provide decision support on a higher level, suggesting complete construction strategies. This is in contrast to the problem addressed here, where we aim at providing detailed information on how to adjust control parameters of the TBM in real time. In both cases, the methods require information about geologic ground class profiles.\\

\noindent\textbf{Geology Prediction}

Geology prediction is an integral part of uncertainty management and has a high influence on real-time decisions. Drilling prior to the start of the project allows one to gather data on the type of geologies that will be met along the trajectory of the tunnel. Depending on the ground condition, it is not always sufficient to gather precise and continuous data on the ground conditions and on the geology changes. A recent study \citecustom{Pawlowski2019}, where several machine learning algorithms for detecting the changes in geology based on the TBM operational data were developed, highlights the difficulty of detecting geology changes from TBM sensor data. This difficulty arises from the fact that the effects of the geology on the TBM performances are intertwined with the control parameter choices of the operators. The discrimination from TBM sensor data between true geology changes and control parameter changes remains an unsolved problem.

\citet{Zhao2019} presented a complete framework for applying a geology type predictor based on TBM operating data consisting of 72 features using a feedforward neural network. In this case study, 88 samples were taken from one single tunnel at 30 meters depth and only operating data recorded in 0.3 m proximity of each sample was considered. This procedure ensures reliable ground truth labels for the approximately 20 geology types, each specified by 7 physical-mechanical indices including natural severity, internal friction angle, deformation modulus, Poisson's ratio etc. 70\% of the data is used for training the model, while 30\% is used for testing. In this case, the model tries to predict all of the 7 indices. For performance measurement purposes, the averaged mean squared error across all indices are reported, in the best case reaching as low as 0.212. Nonetheless, the large amount of samples taken as well as their detailed description reduces the applicability for other tunnel construction projects due to high costs.

Given the difficulty of extracting precise online information on the geology, we decided in our project to separate the geologies into three broad classes. For each geology class, we train a different model, robust to minor geological variations. In real applications, we expect the operators to be able to know broadly the geological type in which the TBM is currently operated (\eg soft, weathered or hard materials).

\noindent\textbf{DSS for Oil Well Drilling}

Decision Support Systems have also been studied in closely related fields such as well drilling \citet{Payette2015}.
With the aim of achieving consistently better drives, a real-time DSS was developed to identify and avoid performance limiters such as bottom-hole assembly whirl, bit balling, stick-slip etc. The proposed framework is an iterative process. One of the core modules of the framework is a learning or calibration phase, in which the operator is encouraged to test different control parameters settings. This procedure enables the system to find optimal regions within the control parameter state-space spanned by the rate of penetration (ROP) and the weight on bit. This approach is not transferable to TBM operation since active parameter exploration is not part of the usual tunnel construction workflow. Similar parameter exploration, particularly for new ground conditions, could be potentially also explored in tunnelling. However, due to the high variability of the ground conditions in long tunnel projects, these explorations would need to be performed in each of the ground conditions, which is not feasible in practical applications. 

A similar approach was presented by \citet{Chandra2019} where deep learning methods and optimization algorithms were applied to generate optimal control parameter recommendations for oil well drilling. More concretely, a feedforward neural network was trained on historical operational data to learn the mapping between the sensor parameters and the rate of penetration. In a second step, recommendations were generated by selecting the control parameters, given by weight on bit, revolutions per minute and flow rate in the pumps, which lead to highest predicted ROP. This selection procedure is based on a heuristic search algorithm, the particle swarm optimization. It requires, therefore, a large number of samples to learn from, to explore the different parameter combinations. It may, therefore, require a simulator if abundant data are not available.

\section{Essential Concepts and Practices of TBM Operation}
\label{sec:essental_conc_for_tbm_op}

To control the TBM during excavation, various control parameters are available to machine operators, making the space of possible setpoint choices very large. The operator's main objective is to achieve safe and efficient operation with the highest possible advance rates. To do so, operators typically adjust the control parameters based on experience and on the expected geology profile. While experienced operators are often able to find a good set of parameters given a certain geology, less experienced operators may, however, operate in sub-optimal conditions. The knowledge of the ground conditions is in any case a centerpiece information for operators to decide on their control actions. In the case study used in this research, operators were only given information on the geology through 13 boreholes drilled over more than half a kilometer, prior to the start of the tunnelling project. Analysis of the project revealed, however, that ground conditions had a high variability and that the samples were not always representative of the real geology profile. Interviews with field experts uncovered that operators decide at the beginning of the excavation phase on the control settings, based on the geology reports from the exploration phase and on their experience. Once the excavation process has started, operators hold their control settings constant as long as possible until it becomes obvious that a sub-optimal state has been reached and that a new control parameter choice is required. Minor parameter adjustments may become necessary to retain the working pressure below the operational limit of the machine. Apart of these adjustments, the operators do, usually, not attempt to optimize the control settings. 

TBMs are commonly equipped with dozens to hundreds of sensors to monitor the state of the system, the environmental and operating conditions while drilling. In this study, real operational sensor data from a TBM were categorized disjointedly into three groups: control parameters (CoP), target parameters (TP) and context parameters (CxP). CoPs relate to the parameters which are directly modified by the operator during excavation to control the TBM. TPs are the parameters to optimise (\eg fastest advance rate, low working pressure). TPs are used to design the optimality function. CxPs represent all remaining sensors capturing essential operating and environmental conditions.\\


\section{Methodology}
\label{sec:meth}

\subsection{Assumptions}
Although the framework, as described in Figure \ref{fig:framework}, was developed based on the case study from a micro-tunnelling machine, the proposed framework can be extended to any mechanized tunnelling system if the assumptions listed below are fulfilled. For the purpose of clarity, the assumptions are sub-divided by their importance into ``imperative'' assumptions, assumptions that have to be fulfilled and ``supplementary'' assumptions that are not crucial for the applicability of the framework but contribute to the performance of the framework.

\noindent\textbf{Imperative Requirements on Data Availability and Quality}

Since the proposed DSS framework is a data-driven framework, the availability of a representative dataset capturing the state of the machine and its interaction with the surrounding ground is required. Commonly monitored parameters include advance rate, thrust, feed pump pressure and cutter head torque. Typically, the system parameters are recorded as time series with defined sampling frequencies. We assume that all sensor time series used to generate recommendations are free from measurement errors. In our case study, field experts selected the most important available parameters and analyzed them in terms of noise and general validity. Furthermore, we assume that the training and testing samples were drawn from approximately the same underlying distribution, that is, from similar machines in similar ground conditions. This is an essential condition for the successful application of data-driven algorithms. One of the possibilities how this condition can be fulfilled is if several tunnels are excavated in close proximity. This was so for the considered case study. A further possibility to fulfil this condition is by selecting similar operating conditions from previous tunnelling projects. Furthermore, the models can also be trained on data collected at the beginning of the project and applied to the subsequent operation (assuming that a sufficiently representative dataset can be collected at the beginning of the project). This assumption would for example be violated if the TBM cutter head shapes or ground conditions between training and testing phases differ substantially.

\noindent\textbf{Supplementary Requirements on Data Availability and Quality}

The supplementary requirements are less crucial. However, fulfilling them can increase the performance of the proposed framework and could help with the interpretation of the results. 

First, since the geology is the most important feature for setting of the control parameters, we assume that the geology profile of the tunnel is available and has a low uncertainty. The geology profile may be provided either by conducting a detailed site investigation prior to tunnelling or by having geologists analyzing samples of excavated materials at the separation unit after drilling. A further option to provide information on geology may be to develop an accurate geology prediction model. However, this model needs to be either developed on previous tunnelling projects with similar geology profiles or on a portion of the data from the current tunnelling project for which the DSS is developed. In our case study, the available geology profiles were rather imprecise. This is why only data collected in homogeneous sections were used for training and testing. 

Second, in order to have a reliable validation reference, the operator control actions (adjustments of CoPs) are assumed to have been recorded and are available for the model development. If this assumption is not satisfied, the control actions can still be reconstructed approximately from the recorded data as introduced in Subsection \ref{subsec:overv_and_preproc}. 

Third, as to ensure robust training and validation we assume to have a large amount of samples available that capture control parameter changes by the operator since only these can be used for the validation procedure introduced in this study. Nonetheless this is usually hard to get since TBM operators tend to rarely change control parameters. 

\subsection{Rationale for the proposed framework}
\label{subsec:overview}

The proposed framework relies on similar concepts as imitation learning~\cite{argall2009survey}. However, it differs from other works in the literature in two ways. First, we introduce a new optimality score to quantify optimal operations. An optimality function is otherwise hard to identify due to the boundary conditions, the operational constraints and the environmental constraints. Second, our framework is designed to mitigate the inherent scarcity of the training data. This scarcity prevents direct application of imitation learning, since the historic operators' actions are only a very small subset of all possible actions and are likely to be sub-optimal. This scarcity of the training data also challenges the quality of the recommendations and this is why we introduce in addition a credibility score. It enables the decision makers to understand the certainty of the applied algorithm under the specific operating conditions.

From the methodological perspective, the proposed framework has three main steps (steps 2.1-2.3 in Figure \ref{fig:framework}), comprising the steps of optimality definition (2.1), training the deep neural network to predict the defined optimality score (2.2) and finally, provide recommendations for the adjustments of the control parameters based on the first order derivative with respect to the control parameters (2.3). Finally, a credibility score is provided to support the decision makers on how certain the algorithms' recommendations are. 

\subsection{Optimality Definition (Fig~ \ref{fig:framework} - 2.1)}
\label{subsec:opt_def}
As discussed above, the goal of the operators is to advance fast and safely through the tunnel trajectory. The operation is, however, constrained by bounds and safety thresholds of the operating parameters. In our case study, if the working pressure (pressure which drives the torque of the cutterhead) exceeds a given safety threshold, the TBM will be shut down automatically, leading to undesired time delays. Based on interviews with tunneling engineers, it appears that the working pressure often changes abruptly, in particular during geological transitions toward harder ground. The operators aim, thus, at maintaining a sufficient margin between the working pressure and the safety threshold to avoid unexpected shut downs. 
Therefore, we propose to define the optimality function $f^{opt}_{\\GC}$ as a continuous linear function, first, of the advance rate, with a positive coefficient to reward higher advance rates, and second, of the working pressure, with a negative coefficient to penalise higher working pressure. To model the concept of safety margin, we propose to increase the working pressure coefficients once it exceeds a certain margin bound ($MB$). The aim is to penalise more heavily the working pressure, once it exceeds this margin.
Since we advocate for training a different model per ground type, we can actually adapt the optimality scoring function to each ground type, enabling realistic expectations of advance rates given the geology. Formally, we propose to define arbitrarily the optimality function for the i-th ground class as
\begin{equation}
f^{opt}_{GC_i}(t) = \left\lbrace
\begin{array}{rl}
\frac{AR_t}{MAR}-w_1\cdot\frac{WP_t}{UB}, &  \mbox{if\ }WP_t\leq MB_i \\[1.2ex]
\frac{AR_t}{MAR}-w_1\cdot\frac{MB_i}{UB}-w_2\cdot\frac{WP_t-MB_i}{UB},     & \mbox{otherwise}.
\end{array}
\right.
\end{equation}
where
\begin{itemize}
\setlength\itemsep{-0.2em}
    \item $\mbox{\small AR}_t$ is the advance rate [mm/min] at time t 
    \item $\mbox{\small WP}_t$ is the working pressure [bar] at time t 
    \item $\mbox{\small UB}$ is the upper bound of the working pressure (safety threshold before automatic shutdown)
    \item $\mbox{\small MB}_i$ is the working pressure margin bound [bar] defined as the observed 90th-percentile for the i-th ground class
    \item $\mbox{\small MAR}_i$ is the observed maximum advance rate within i-th ground class (\GC)
    \item $\mbox{\small w}_1$ is the negative penalizing weight on the working pressure, when the working pressure is below the margin bound $\mbox{\small MB}_i$.
    \item $\mbox{\small w}_2$ is the negative penalizing weight on the working pressure, when the working pressure is above the margin bound $\mbox{\small MB}_i$. Typically we have $w_2\gg w_1$.
\end{itemize}
The values of the hyper-parameters, notably the penalizing weights on the working pressure $\mbox{\small w}_1$ and $\mbox{\small w}_2$, can either be decided \textit{a priori} based on expert knowledge, or tuned \textit{a posteriori} based on the performance of the model in test conditions.

For the purpose of simplicity, we only consider the advance rate and the working pressure in our definition of the optimality. However, other constraints or parameters could be integrated if deemed relevant by the TBM operator. The proposed methodology can be easily applied to any other optimality definition. 

Also, to ease the understanding by the operator, we propose to normalise the optimality score as a value between 0 and 100, where 100 corresponds to highest achieved score observed in historic data.

\subsection{Training: Learn Mapping from TBM State to Optimality (Fig~ \ref{fig:framework} - 2.2)}
\label{subsec:training}
In the second part of the framework, a machine learning model is developed and trained to learn the mapping from the control parameters (CoP) and the context parameters (CxP) to the previously defined optimality function $f^{opt}_{\GC_i}$. In this step, having historic data with a sufficiently rich representation of the different parameter combinations is crucial for training a model able to accurately predict the optimality given varied CoP and CxP and to provide good recommendations. To perform this regression task, we propose to use a deep feed forward neural network, as it has shown excellent performance in complex and varied regression tasks.(\citet{Alex2012}, \citet{Szegedy2014}).
To select the hyperparameters of the neural network, we follow here standard validation practices such as k-fold cross validation combined with a grid search. The process is detailed in Section~\ref{subsec:grad_mod}.

\subsection{Generating Recommendations and Credibility (Fig~ \ref{fig:framework} - 2.3)}
\label{subsec:recommend}

In the previous section, we proposed to train a machine learning model which, given current control parameters and context parameters, is able to estimate the optimality score. In this step, we propose to compute the first order derivative of the model's output with respect to the control parameters, that is, to compute which modifications of the control parameters current value lead to an increased optimality score. This allows us to recommend in real-time at each consecutive time step one incremental modification of the control parameters toward higher scores of the optimality. 

Since the validity of the recommendation highly depends on the quality of the model and, therefore, on its training dataset and its representativeness, we propose, additionally, to compute a credibility measure to quantify the recommendations' trustworthiness. In this work, we propose to rate the recommendation based on the local performance of the model on similar controlled and context parameters in historic data (the nearest neighbours).
For each of the $n$ nearest neighbours, we evaluate the performance of the model with two combined metrics. First, we evaluate the quality of the estimation of the optimality score in the neighbourhood of the current observation to quantify the performance of the learning. To do so, we evaluate the $\ell_2$-norm of the difference between the predicted optimality and the true optimality for the j-th nearest neighbours ($e^j_1$). Second, we evaluate the recommendations of the model with similar points from historic data by comparing the recommendations and the actions taken. That is, for the j-th nearest neighbour and the consecutive observation, we compute the $\ell_2$-norm of the difference between the computed otpimality gradient and the observed optimality difference ($e^j_2$). To combine these metrics, we first normalise them based on typical values observed in the validation set (using the 5-th and 95-th percentile). We then average the two metrics and compute their complement to 1, such that higher values indicate a higher credibility. Formally, we perform the following operations:
\begin{equation}
\forall i\in \lbrace1,2\rbrace,\quad e^j_{norm,i} = clip(\frac{e_i - Q^{i}_5}{Q^{i}_{95} - Q^{i}_5},0,1),
\end{equation}
\begin{equation}
T^j = 1-\frac{e^j_{norm,1}+e_{norm,2}}{2},
\end{equation}
where $clip(z,0,1)$ is the clipping function in $[0,1]$. 
$Q^{i}_{95}$ and $Q^{i}_5$ are the 95-th and 5-th percentiles of the i-th metric on the validation set. 
 
The final credibility of a recommendation, for the $k$-th sample, is the Gaussian-weighted mean of the trust values $T^j$ over its $n$ nearest-neighbours set, $\mathcal{N}(k)$. That is:
\begin{equation}
   \mbox{Credibility}_{k} = \frac{1}{n} \cdot \sum_{j\in\mathcal{N}(k)} w^k_j \cdot T^j,
\end{equation}
with
\begin{equation}
\large
    w^k_j = \exp{\left( -\frac{\lVert\mbox{CxP}_{k} - \mbox{CxP}_j\rVert_2^2}{B^2}\right)}
\end{equation}

To choose the number of neighbours to consider, $n$, we computed in historic data the average distance to the k-th nearest neighbour as a function of $k$. We found an abrupt change of slope when 15 neighbours are considered and decided therefore to set $n=15$. The Gaussian kernel width parameter $B$ is chosen as the average standard deviation of the distances to the $15$ nearest neighbours.

\section{Case Study}
\label{sec:case}

\subsection{Project Overview}
\label{subsec:proj_overv}

The rapid increase of renewable energy in the northern part of Germany has led to over capacities, which needed to be reduced by transmitting the excess energy to the southern part of the country. Transmission grids are usually constructed overground and must satisfy many requirements posed by local residents and property owners, which often leads to cumbersome and prolonged procedures. For this reason, the government has decided to support the expansion of the grids underground, making use of the microtunnelling technology to circumvent such problems. One of these projects was completed between 2018 and 2019. The TBM used in this project was designed for pipe jacking. It has a small diameter of about half a meter and is designed for fast and efficient drilling procedures. The TBM was fully equipped with dozens of sensors sampling at 0.1Hz. A total of six micro-tunnels with a length 688 meters each were excavated \citecustom{epowerpipe}.

\subsection{Data Overview and Pre-processing}
\label{subsec:overv_and_preproc}

\noindent\textbf{Sensor Data}

Although 6 micro-tunnels were excavated, an initial analysis of the data led us to restrict the analysis to two of the micro-tunnels. The other micro-tunnel data were left aside in this pilot study since some had missing geological information and other had different data collection systems. Furthermore, based on discussions with domain experts, 26 relevant parameters were selected for the analysis. As described in Section~\ref{sec:essental_conc_for_tbm_op}, we split the parameters into 3 groups: the 2 operating parameters used to compute the optimality score (advance rate and working pressure), the 5 parameters corresponding to the parameters controlled by the operator (CoP) and 19 context parameters (CxP). The CoP are: the rotational speed of the cutter head ($\CoP_1$), often set at its maximum value to achieve full excavation rate; the high pressure water nozzle ($\CoP_2$), used to clean the cutting wheel and ensure efficient excavation; the drive line pressure ($\CoP_3$), used to control the vacuum created to transport the material rapidly to the separation unit; the jacking frame thrust ($\CoP_4$), controlling the load applied at the cutting wheel and the feed pump rotational speed ($\CoP_5$), controlling the amount of water/slurry used to extract the material into the excavation chamber.
In this case study, the true control actions taken by the operator on the CoP were not directly recorded. Therefore, the above CoP are inferred from closely related sensor measurements. We observed that these related measurements had distinct small and large variations, due respectively to noise and operator actions. External causes, such as sensor noise and ground interactions will lead to small fluctuations of the measurements while operator actions will have larger impact on the sensor values. In fact, when looking at the distribution of these fluctuations, one can observe two modes, one for small fluctuations and one for larger fluctuations, with a drop in-between. This drop can be used as a threshold to recover the operators' actions.

The context parameters, CxP, listed in Table~\ref{tab:context_params}, are a combination of pressure-, temperature-, force-, flow rate- and positional parameters captured by sensors installed across the TBM.\\

\begin{table}
\centering
\renewcommand{\arraystretch}{0.7}
\caption{A list of all context parameters \label{tab:context_params}}
\begin{tabular}{lc}
     \toprule
     \multicolumn{1}{c}{\textbf{Context Parameters}}\\
     \midrule
      Pressures [bar] \\
      \midrule
      Steering cylinder 1 pressure\\
      Steering cylinder 2 pressure \\
      Steering cylinder 3 pressure\\
      Steering cylinder 3 pressure (3-B)\\
      Feed line pressure on TBM\\
      Feed line pressure on pump\\
      Suction line pressure \\
      Bentonite pump pressure \\
     \midrule
      Flow rates [$m^3$/s] \\
      \midrule
      Conveyor line flow rate\\
      Feed line flow rate\\
      Drive line flow rate\\
      High pressure nozzle flow rate \\
     \midrule
      Other\\
      \midrule
      High pressure pump rotational speed [rpm]\\
      Bentonite pump rotational speed [rpm]\\
      Steering cylinder 1 extension [mm]\\
      Steering cylinder 2 extension [mm]\\
      Steering cylinder 3 extension [mm]\\
      Machine oil temperature [celsius]\\
      TBM axial rotation [degrees]\\
    \bottomrule
    \end{tabular}
\end{table}

\noindent\textbf{Geology Data}

Optimal operational parameters depend on the geology surrounding the TBM. Yet, the geological information is hard to obtain for two reasons: first, predicting the surrounding geology from sensor data remains an open research question, second, studies prior to the start of the project only collect scarce samples and contain many uncertainties. In this project, however, geologists could analyse the excavated material and provide \textit{a posteriori} the geology profile over the entire tunnel trajectory. Nevertheless, this profile contains some uncertainties since part of the analysed excavated material is dissolved in the slurry. Based on discussions with domain experts, we decided to consider three geological classes, for which uncertainties were deemed low enough by the experts and for which sufficiently long homogeneous sections could be identified. Based on operational data collected in these homogeneous geologic profiles, we trained one independent model for each of the geology classes. In real applications, we expect that the operator is able to know roughly the type of geology in which the TBM is currently operated to choose the right model. The three considered geology classes are:
\begin{itemize}
    \item GC1: Homogeneous highly weathered schist (soft)
    \item GC2: Homogeneous moderately weathered schist (firm)
    \item GC3: Homogeneous slightly weathered schist (hard)
\end{itemize}

\noindent\textbf{Data Pre-processing}

In order to prepare the dataset for the case study, we performed the following operations on the raw data:
\begin{enumerate}
    \item Data Cleansing
    \begin{enumerate}
        \item Removal of TBM retraction phases (\eg for TBM cutting wheel maintenance)
        \item Removal of sections in which the tunnel length decreases or is constant, due to anomalous measurements or to actions other than the excavation.
        \item Removal of unrealistic/erroneous sections (after validation with a field expert), including sections with unrealistic values for the parameters.
        \item Removal of start-up and shut-down transient phases of the TBM \citecustom{Zhang2019}.
    \end{enumerate}
    \item Smoothing of all parameters with a Gaussian Kernel Smoothing Algorithm with a bandwidth of 30 seconds (3 samples).
    \item Standardization of all features (removal of their mean and scaling by their standard deviation)
\end{enumerate}

\subsection{Optimality Assessment}
\label{subsec:opt_ass}
The parameters determining the proposed optimality function as defined in Section \ref{subsec:opt_def} are given by 

\begin{itemize}
    \item the slope parameters $\mbox{\small w}_1=0.8$ and $\mbox{\small w}_2=3.0$,
    \item the margin bound on the working pressure for each ground class with 114 bar for GC1, 124 bar for GC2 and 126 bar for GC3.
\end{itemize}

\subsection{Baseline Model (Nearest Neighbors)}
\label{subsec:baseline}
To evaluate the benefits of our approach, we propose to compare the developed framework to a simpler approach: finding in historic observations points with similar context parameters (CxP) but higher optimality score and using their control parameters (CoP) as recommendation. The idea for this approach is based on the concept that if two observations have similar CxP, but one of the observations has a higher optimality score, the differences can likely be explained by better CoP.

For this baseline, we use a nearest neighbor algorithm (\citet{Cover1967},\citet{Chen2018}).
Similarly as described above, we set the numbers of considered neighbours to 15, and propose to recommend for the $k$-th sample, the Gaussian-weighed average of the CoP of its neighbour set, $\mathcal{N}(k)$. Formally this is computed as:
\begin{equation}
\large
    \mbox{R}^{nn}_{k} =  \frac{1}{n} \cdot \sum_{j\in\mathcal{N}(k)}  w^k_j \cdot (\mbox{CoP}_{j} - \mbox{CoP}_{k})
\end{equation}

\subsection{Gradient-Based Recommendations}
\label{subsec:grad_mod}
\begin{figure}
\setlength{\tabcolsep}{1pt}
\begin{tabular}{cc}
(a)&\includegraphics[align=c,width=12cm]{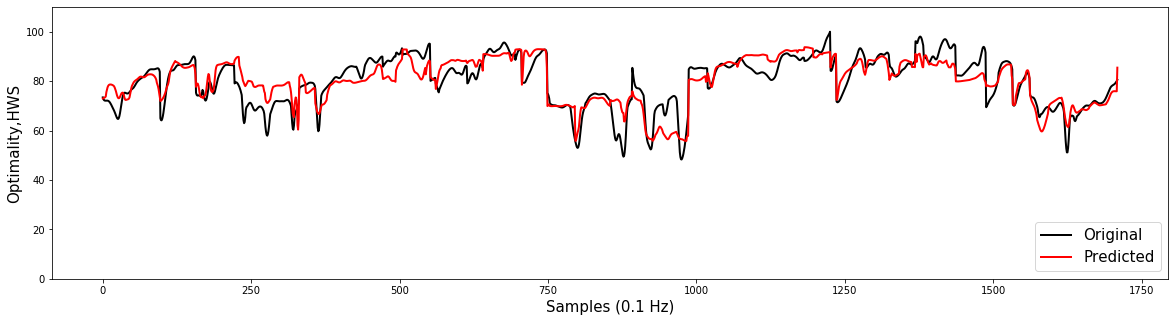}\\ 
(b)&\includegraphics[align=c,width=12cm]{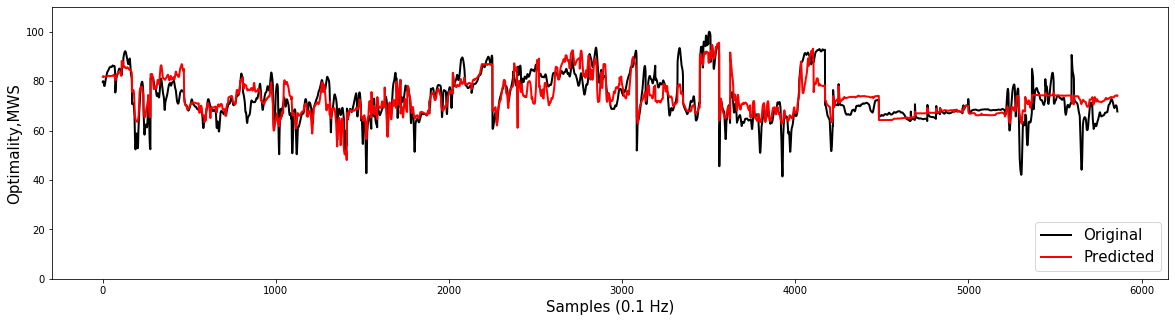}\\
(c)&\includegraphics[align=c,width=12cm]{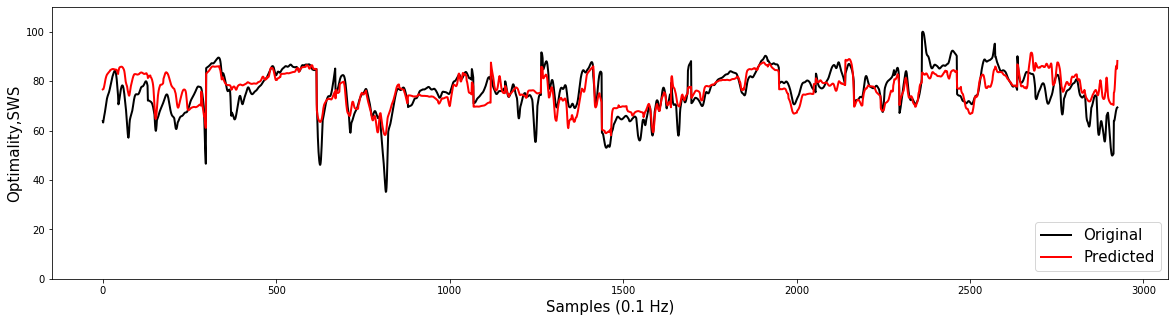}\\
\end{tabular}
\caption{\textbf{Predicted vs. true optimality for (a) \GC$_1$, (b) \GC$_2$ and (c) \GC$_3$}\label{fig:optimality_pred_sws}}
\end{figure}

The neural networks (\cf Section \ref{subsec:recommend}) used in this study to learn the relationships between CxP, CoP, and the optimality score were designed in the following way.
We performed a grid search on the neural network for the first ground class GC1 with the following parameter ranges: the number of layers (from 2 to 4), the number of neurons (from 30 to 1000 in steps of 10), the dropout rate (from 0 to 0.3 in steps of 0.1) and the learning rate (from $10^{-4}$ to $10^{-1}$ in steps of factor 10).
We observed that training the network over 200 epochs led to the convergence of the training. Using 10-fold validation, we found that the architecture summarised in Table~\ref{tab:model_params} minimises the $\ell_2$-norm of the difference between the estimated and the ground truth optimality score. This architecture was subsequently used for all neural networks in this work. For illustration purposes, examples of learned versus ground truth optimality scores for each ground class are shown in Figure~\ref{fig:optimality_pred_sws}.

\begin{table}
\centering
\renewcommand{\arraystretch}{0.7}
\caption{Hyper-parameters for the applied feed forward neural network\label{tab:model_params}}
\begin{tabular}{lc}
     \toprule
     \multicolumn{1}{c}{\textbf{Parameters of the FNN}}\\
     \midrule
      Architecture \\
      \midrule
      Number of layers : 4\\
      Number of neurons in layer [1-3]: 50 \\
      Number of neurons in last layer: 1\\
      Dropout rate: 0.2 \\
      Activation function: sigmoid\\
     
     \midrule
      Optimizer \\
      \midrule
      
       Adam\\
       Learning rate: 0.01\\
       Decay rate: 0.9\\
       Loss: L2 \\

     \midrule
      Additional\\
      \midrule
      Number of epochs: 200\\
      Batch size: 200\\
      
    \bottomrule
    \end{tabular}
\end{table}

\subsection{Validation Methods}
\label{subsec:valid_meth}
The validation of a recommendation framework in a non-reproducible environment is always challenging if a simulation environment is not available. Indeed, one cannot bore the same tunnel more than once to test different scenarios and the effect of different CoP combinations.

The validation of the proposed framework under real conditions will require that the system is applied in real time to evaluate the operators' actions and to perform a statistical analysis on the overall performance of several operators with and without recommendations (under similar operating conditions). 

Since such large-scale experiments are difficult to implement, we propose a surrogate validation method. In our case, only historic data are available for validation and due to the multi-dimensionality of the CoP, different actions might lead to similar effects on the optimality score. A difference between an operator's action and a recommendation cannot be used to invalidate the recommendation. Thus, we propose to evaluate instead the cases where the operators' actions and the recommendations agree. For each CoP, we compute the ratio of cases where the change in optimality score predicted by our model matches the observations. We compute these ratios in two ways. First, we compute this ratio on historic data. This is a reliable indicator but hard to estimate since often, the operator actions and the recommendations do not match. We denote this ratio as ``Synchronised Validation''. Second, to account for the few cases where both the operators' actions and the recommendations match, we propose to find other similar samples in historic data to compare with. Within the samples with similar Context Parameters, identified with the nearest neighbours model, we compare to the sample with the most similar CoP and a different optimality score. We denote this score as ``Contextual Validation''. This second approach also allows us to consider control parameters from other operators during other parts of the project as reference points. Both methods to compute the validation indicators are described in Algorithm~\ref{alg:Val}.

\begin{algorithm}[!htbp]
\linespread{1.2}
	\caption{Validation}
	\label{alg:Val}
	\begin{algorithmic}[1]
	\Input $\CoP_i, \CxP_i, \forall i \in \llbracket1..5\rrbracket$
	\State $ValSV_i \gets 0, NumSV_i\gets 0 \quad \forall i \in \llbracket1,5\rrbracket$
	\State $ValCV_i \gets 0, NumCV_i\gets 0 \quad \forall i \in \llbracket1,5\rrbracket$
	\For{all $t$ timestep available}
    	\State $\GC_t \gets$ ground class of sample $t$
    	\State $\widehat{\CoP}^{t+1} = Recommendation(\CxP^t, \CoP^t,\GC_t)$
    	\For{$i\in \llbracket 1,5 \rrbracket$}
    	\If{$sign(\widehat{\CoP}_i^{t+1}) == sign(\CoP_i^{t+1})$}
    	   \State $NumSV_i += 1$
    	   \If{$sign(f^{opt}_{\GC_t}(t+1)) == +1$}
    	      \State $ValSV_i += 1$
    	      \EndIf
    	   \EndIf
    	\EndFor
    	\State \ 
    	\State $\mathcal{N}_t \gets$ 15-closest neighbour set of $\CxP^{t}$
    	\State $t' \gets \Argmin{\hat{t}\in \mathcal{N}_t}{\Vert \widehat{\CoP}^t - \CoP^{\hat{t}}\Vert_2}$
    	\For{$i\in \llbracket 1,5 \rrbracket$}
    	\If{$sign(\widehat{\CoP}_i^{t+1}) == sign(\CoP_i^{t'})$}
    	   \State $NumCV_i += 1$
    	   \If{$sign(f^{opt}_{GC_t}(t')) == +1$}
    	      \State $ValCV_i += 1$
    	      \EndIf
    	   \EndIf
    	\EndFor
	\EndFor
\State $SV_i \gets ValSV_i/NumSV_i \quad \forall i \in \llbracket1..5\rrbracket$
\State $CV_i \gets ValCV_i/NumCV_i \quad \forall i \in \llbracket1..5\rrbracket$
\State $SV= \Avg{i \in \llbracket1,5\rrbracket}{SV_i}$
\State $CV= \Avg{i \in \llbracket1,5\rrbracket}{CV_i}$
	\end{algorithmic}
\end{algorithm}

\section{Results}
\label{sec:res}

\begin{table*}[tph]
\centering
\renewcommand{\arraystretch}{0.7}
\caption{\textbf{Validation Scores:} For the two models, Nearest Neighbours (NN) and Gradient-based (GB), the achieved score on the validation set for each control parameter (CoP$_i$ in columns), for each geological class (\GC$_i$ in rows) and for both proposed validation indicators (Synchronized and Contextual). The averages are also reported.\label{tab:scores}}
\hspace*{-2cm}
\resizebox{1.2\textwidth}{!}{
\setlength{\tabcolsep}{4pt}
\begin{tabular}{ l|c@{\hspace{2pt}}c@{\hspace{8pt}}c@{\hspace{2pt}}c@{\hspace{8pt}}c@{\hspace{2pt}}c@{\hspace{8pt}}c@{\hspace{2pt}}c@{\hspace{8pt}}c@{\hspace{2pt}}c|c@{\hspace{2pt}}c||c@{\hspace{2pt}}c@{\hspace{8pt}}c@{\hspace{2pt}}c@{\hspace{8pt}}c@{\hspace{2pt}}c@{\hspace{8pt}}c@{\hspace{2pt}}c@{\hspace{8pt}}c@{\hspace{2pt}}c|c@{\hspace{2pt}}c}
     \toprule
     & \multicolumn{12}{c||}{\textbf{Synchronized Validation}} & \multicolumn{12}{c}{\textbf{Contextual Validation}}\\
     \midrule
      \GC &  \multicolumn{2}{c}{$\CoP_1$} & \multicolumn{2}{c}{$\CoP_2$} & \multicolumn{2}{c}{$\CoP_3$} & \multicolumn{2}{c}{$\CoP_4$} & \multicolumn{2}{c|}{$\CoP_5$} & \multicolumn{2}{c||}{$Avg$} &  \multicolumn{2}{c}{$\CoP_1$} & \multicolumn{2}{c}{$\CoP_2$} & \multicolumn{2}{c}{$\CoP_3$} & \multicolumn{2}{c}{$\CoP_4$} & \multicolumn{2}{c|}{$\CoP_5$} & \multicolumn{2}{c}{$Avg$} \\
      \midrule
      & NN & GB & NN & GB& NN & GB& NN & GB& NN & GB& NN & GB& NN & GB& NN & GB& NN & GB& NN & GB& NN & GB& NN & GB\\
      \midrule
      
      \GC$_1$ & 58 & 80 & 56 & 51 & 72 & 50 & 60 & 68 & 63 & 59 & 62 & 62 &
      43 & 55 & 40 & 62 & 52 & 63 & 61 & 57 & 53 & 75 & 50 & 63\\
      \GC$_2$ & 63 & 80 & 51 & 50 & 48 & 51 & 50 & 73 & 55 & 59 & 54 & 63 &
      50 & 54 & 54 & 48 & 54 & 51 & 53 & 49 & 45 & 52 & 51 & 51\\
      \GC$_3$ & 73 & 85 & 52 & 51 & 53 & 48 & 59 & 67 & 57 & 62 & 59 & 62 & 
      51 & 44 & 28 & 42 & 60 & 43 & 58 & 43 & 53 & 43 & 50 & 43
      \\
      
     \midrule
     $Avg$ & 65 & 82 & 53 & 51 & 58 & 50 & 56 & 69 & 58 & 60 & \textbf{58} & \textbf{62} & 
     48 & 51 & 41 & 51 & 55 & 52 & 57 & 50 & 50&57 & \textbf{50} & \textbf{52}\\
    \bottomrule
    \end{tabular}}
\end{table*}

\begin{figure}
\centering
\includegraphics[width=8cm]{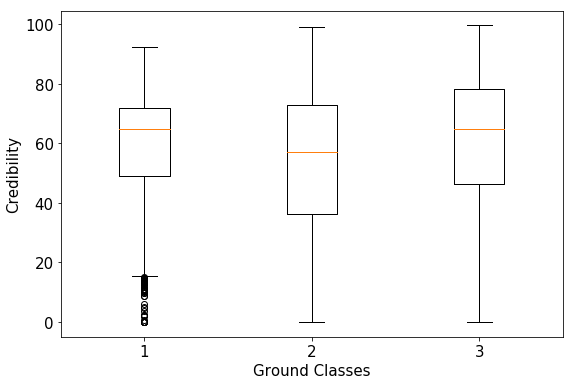} 
\caption{\textbf{Distributions of the Credibility Scores computed for the three geology ground classes ($GC_1-GC_3$)}\label{fig:cred_distr}}
\end{figure}

\begin{figure*}
\centering
\includegraphics[width=1\textwidth]{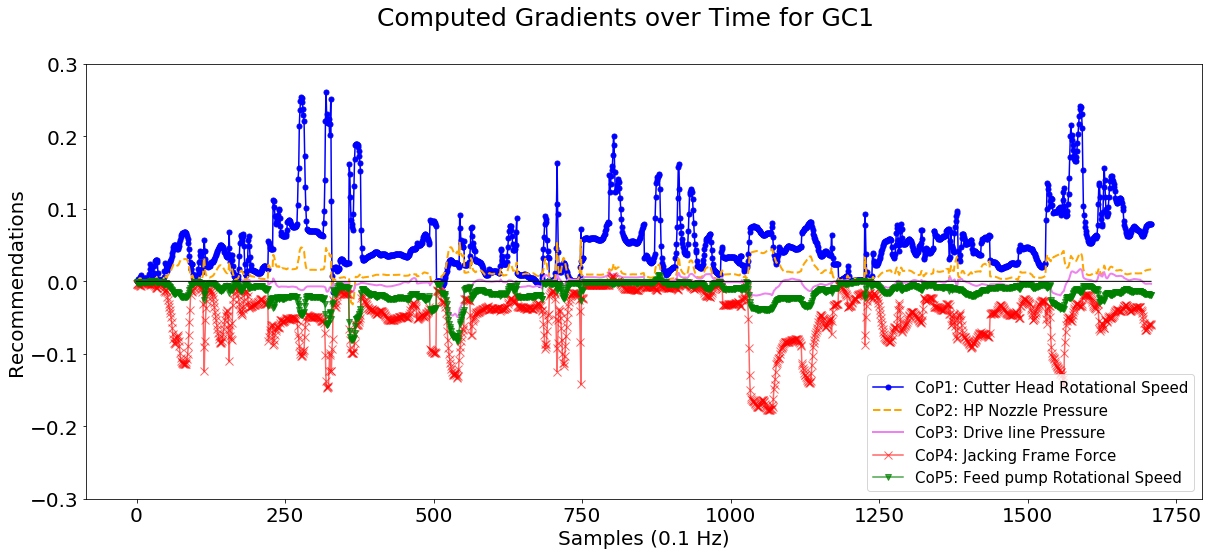}
\caption{\textbf{Gradients over Time for GC1}\label{fig:gradients}}
\end{figure*}

Both validation scores, for the proposed gradient-based (GB) recommendations and the applied nearest-neighbour baseline (NN) are presented per CoP and per Ground Class in Table~\ref{tab:scores}.

Overall, the gradient-based model achieves higher agreement with historic observations, demonstrating the value of the proposed framework. 

The ``Contextual Validation'' (CV) scores are lower on average compared to the ``Synchronised Validation'' (SV). This can have two explanations. First, it can be that some contextual information is not captured by any sensors. This could lead to an hidden discrepancy between neighbouring points. Second, it could also be explained by a lack of sufficiently similar samples in the dataset.

The Synchronised Validation scores in Table~\ref{tab:scores} are fluctuating less for the NN model than for the GB model: the ranges are 58-63 versus 50-80, 48-63 vs. 50-80 and 52-73 vs. 51-85 for $GC_1$, $GC_2$ and $GC_3$ respectively. This is expected since the NN model inherently reproduces historic operators actions over all parameters. However, a discussion with the experts validated that the recommendation of the gradient model are very consistent with operators' knowledge on the system. In particular, the gradient model achieves much higher scores for $\CoP_1$ and $\CoP_4$, which are coincidentally the two parameters for which the model computes the highest gradients as illustrated in Figure~\ref{fig:gradients}. The gradient for $\CoP_1$ is mostly positive (the model advises to increase the cutter head rotational speed), which corresponds to the expert feedback that this parameter is the one that operators attempt to maximise. In fact, since thresholds for CoP were not included in the current model, part of the cases where the recommendation and the observations do not agree could be explained by cases where this parameter is already at its maximum setting. In Fig.~\ref{fig:gradients}, the gradient for $\CoP_4$ is mostly negative, meaning that the model mostly advises for a decrease of the jacking frame force. This is an interesting insight since higher force should, according to the experts, help with the boring. However, a local change of geology might affect  the working pressure in a  dangerous way if the jacking frame force is too high. Therefore, the model may be minimising this risk, in particular since we designed our optimality score such as to strongly discourage high working pressure.

For the remaining parameters, $\CoP_2$ has mostly a positive gradient in Fig.~\ref{fig:gradients}, that is, the model advises most of the time to increase the nozzle pressure. The experts validated this recommendation since it cleans the head from sticking residuals that lower the excavation capacity. $\CoP_5$ has mostly a negative gradient in Fig.~\ref{fig:gradients}, so the model frequently advises to decrease the feed pump that carries the water into the excavation chamber. According to experts, too much water can increase the difficulty of excavation and lead to lower advance rates. Last, the drive line pressure has a very low gradient magnitude on average. Since the drive line controls the transport of the excavation material to the separation unit, experts say that the setting of this parameter mostly depends on the distance between the head of the TBM and the separation unit and is independent of the advance rate and of the working pressure. A gradient close to zero confirms this observation in Figure~\ref{fig:gradients}.

These experts' insights, both on the sign and on the magnitude of the gradients confirm the relevance of the recommendations. The magnitude of the gradient can also be interpreted as a measure as to which CoP change would lead to the highest optimality score improvement.

Figure~\ref{fig:cred_distr} presents the distribution of the computed ``credibility'' scores over the three ground classes. $GC_1$ has the highest median and lowest inter-quartile range which can be explained by a better learning of the optimality function in the neighbourhood of each sample, and by a denser sample distribution. This is in line with the overall best results achieved by both models on $GC_1$, in particular for the nearest neighbour baseline.

\section{Discussion}
\label{sec:disc}

\subsection{Data Scarcity}
\label{subsec:Data_unav}
Similarly to any real project, accessing a sufficiently large number of samples for training, validation and testing is of primary importance for the proposed framework. In this project, if the original dataset contained 0.1Hz data for six drives or around 1 million samples. We could only use data from two of the micro-tunnels, that is from 275\,000 samples. The other micro-tunnels had large missing gaps, inadequacies within the drilling process or non-availability of the detailed geology profile. For these two micro-tunnels, after the data pre-processing around 70\,000 samples are left, and once we extracted the three homogeneous geologies we are left with around 5\,000 samples per section. Last, for the validation set, we could only use samples with control parameter changes (to compare them to the recommendation of the proposed framework). All of these applied constraints for the selection, left only around 2\,000 samples per homogeneous geology (around 0.6\%). 
We believe that with more data, the gradient approach will improve further and with the systematization of data collection during operation and with the identification of the most important parameters, the issue of the availability of representative training data should be mitigated in the future.

\subsection{Optimality Definition}
\label{subsec:opt}
In this project, the definition of the optimality score was designed in a simple way to favor high but safe tunnel boring speeds. The proposed function can be tuned to improve the model recommendations. For example, if the model recommendations are too conservative (or respectively not sufficiently conservative) regarding the working pressure, one could  lower (resp. increase) the penalizing weights $w_1$ and $w_2$ that have a direct effect on the optimality score. More generally, the same framework could be used with much more complex optimality functions. Such more complex optimality functions could include for example financial factors, component degradation or slurry density increase. Before each new each project, the most relevant optimality function for that specific project respecting the needs of the operators can be designed and the model retrained based on historic data.

\subsection{Unavailable Operator Actions}
\label{subsec:reconstruct}
In this project, we could not access the operator true actions and had to reconstruct them from related measured parameters as explained in Section~\ref{subsec:overv_and_preproc}. This uncertainty on the true operator actions can propagate to the trained model and to our validation methods. Thereby, it could lower the validation scores. In future projects, we recommend to record operator actions as well as measured parameters to reduce these uncertainties.

\subsection{Validation Biases}
\label{subsec:valid}
\begin{figure}
\centering
\includegraphics[width=8cm]{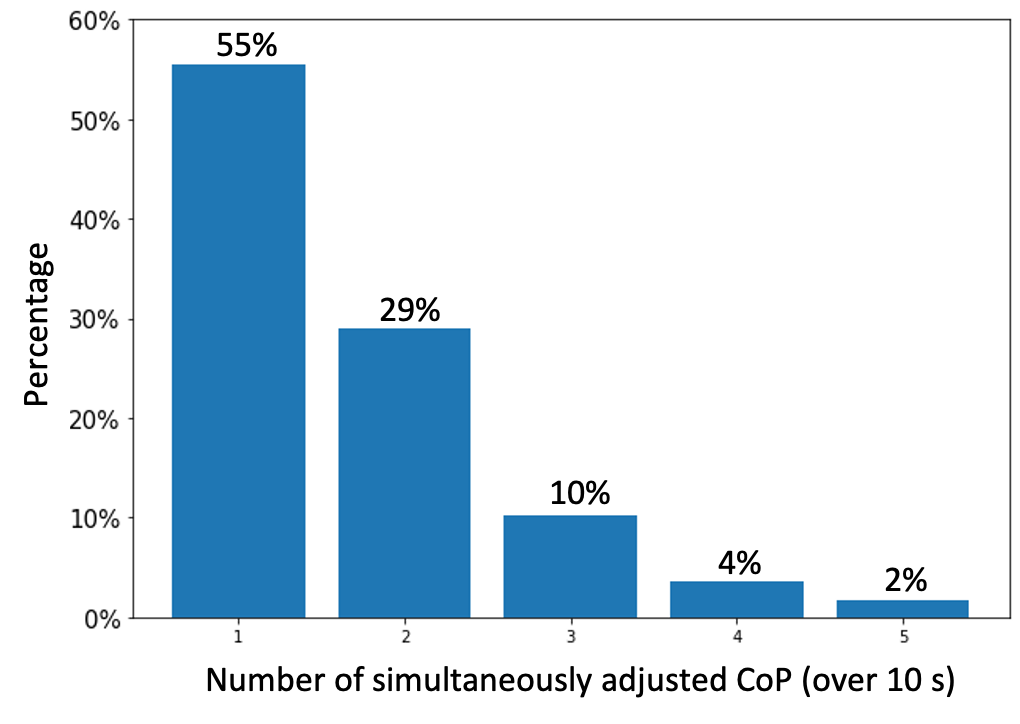}
\caption{\textbf{Frequency of the number of simultaneously adjusted control parameters by the operators over all three ground classes for both selected micro-tunnels.}\label{fig:sp_drive5}}
\end{figure}

The proposed validation methodology evaluates the recommendations on each parameter independently. This is partly motivated by the gradient approach which computes the partial derivative of the optimality score with respect to each control parameter, that is, assuming that other parameters remain constant. Yet, since the CoP are a multi-dimensional space, their combined actions could be considered together. In practice, we observe that the operators are hardly ever changing the control parameters, and when they do, very few parameters are adjusted at the same time, as illustrated in Figure \ref{fig:sp_drive5}. In $55\%$ of the cases, the operators only adjust one CoP. Without a much larger dataset, it would be highly improbable to find validation points for which the 5 CoP were changed simultaneously in a similar way to the recommendations, given similar CxP. This gap between the model interpretation of the control parameters as a multi-dimensional space and the observations that only few dimensions are changed at the same time could also explain lower validation computed on historic data.

\subsection{Credibility Measure}
\label{subsec:cred}

The limitations due to data scarcity and validation bias are captured, to some extend, by the credibility score proposed within this framework. The relatively low observed credibility measures can be linked to the difficulty to find sufficiently representative neighbours for most samples. This lack of relevant neighbours affects the credibility score twice. First, since few data points are available, the model performance in predicting the optimality score will be lower. Second, the credibility score is weighted down by the distance to the neighbours. This score, therefore, plays its role of indicating to which extend the current recommendation is supported by historic data. We expect the overall credibility score to increase as more and more data become available and as the model also learns the optimality score better. Yet, if new conditions are encountered, the credibility will drop, letting the operator know that he or she should proceed with caution.

\subsection{Computational Complexity}
\label{subsec:cc}

As the amount of available representative data increases, we expect the models to perform better. Yet, this will also affect the computational requirements. The nearest neighbour algorithm complexity at testing time is $\mathcal{O}(k\log n)$ where $k$ is the number of neighbours and $n$ the number of samples~\cite{Moore-1991-15799}.
The complexity of the gradient model at testing time is $\mathcal{O}(1)$ and is, therefore, expected to scale without any problem to much larger datasets and be more suitable for real-time applications. The credibility measure also relies on the nearest-neighbour algorithm and scales as $\mathcal{O}(k\log n)$. Yet, as discussed above, we expect that when the dataset becomes sufficiently large, this measure would not be needed anymore. For example, on a normal 3-years old laptop, the Nearest Neighbours model computational time increases from 8.7 to 97.2 seconds when the number of samples increases from $1700$ to $5800$ in the dataset, while the Gradient-based model's time increases only from 2.60 to 2.85 seconds.

\section{Conclusion}
\label{sec:conc}

In this work, we proposed a complete framework for the development of a decision support system for advising TBM operators on how to adjust control parameters in response to rapidly changing conditions. This framework uses the information encoded in sensor data to support the machine operator's control actions directly during the drilling procedure. Thus, it complements existing work to quantify and understand sources of uncertainties and to improve risk management in tunnel construction projects. The proposed framework relies on the definition of an optimality scoring function, which can be designed with expert knowledge as a multi-objective function, such as in our case study, maximising the advance rate while minimising the working pressure. With the defined optimality scoring function, the framework learns from historic data a mapping between environmental conditions, control parameters and the optimality score. It then recommends how to adjust control parameters in order to improve on the optimality score. We proposed in addition a credibility score which analyses the performance of the model on historic data in the neighbourhood of the current conditions. Although the validation of recommendation systems with historic data is challenging, we proposed to evaluate two different matching scores, whose results show consistency between recommendations of the proposed framework and historic data. The recommendations could also be validated and explained by experts, showing high potential for the proposed framework.

Further improvements are to be expected as data will be more and more consistently collected across projects, types of TBM, operators and geological conditions. This could lead to further validation of the framework before its implementation in real conditions. The future development of the framework would greatly benefit from more explorative actions from the operators, which could be encouraged by such recommendation systems. The proposed framework can be implemented in a continuous learning scheme, as data are collected in real time. The natural next step, once a sufficiently robust system will be running and sufficient representative samples are collected, would be to develop the framework further to a reinforcement learning framework. Such approaches are left for future work.

\section{Acknowledgement}
\aknow

\bibliography{sample}

\end{document}